\documentclass{llncs}
\usepackage{makeidx}  %
\usepackage{epsfig}  %
\begin{document}
\frontmatter          %
\mainmatter              %
\title{Session Initiation Protocol (SIP) Server Overload Control: Design and Evaluation}
\titlerunning{SIP Server Overload Control}  %
\author{Charles Shen\inst{1} \and Henning Schulzrinne\inst{1}
 \and Erich Nahum\inst{2}}
\authorrunning{Charles Shen et al.}   %
\tocauthor{Charles Shen, Henning Schulzrinne, Erich Nahum}
\institute{Columbia University, New York, NY 10027, USA\\
\email{\{charles,hgs\}@cs.columbia.edu}%
\and
IBM T.J. Watson Research Center, Hawthorne, NY 10532, USA\\
nahum@watson.ibm.com}
\maketitle              %

\begin{abstract}

A Session Initiation Protocol (SIP) server may be overloaded by emergency-induced call volume,
``American Idol'' style flash crowd effects or denial of service
attacks. The SIP server overload problem is interesting especially
because the costs of serving or rejecting a SIP session can be
similar. For this reason, the built-in SIP
overload control mechanism based on generating rejection messages
cannot prevent the server from entering congestion collapse under
heavy load. The SIP overload problem calls for a pushback control
solution in which the potentially overloaded receiving server may
notify its upstream sending servers to have them send only the
amount of load within the receiving server's processing capacity.
The pushback framework can be achieved by either a rate-based
feedback or a window-based feedback. The centerpiece of the
feedback mechanism is the algorithm used to generate load
regulation information. We propose three new window-based feedback
algorithms and evaluate them together with two existing rate-based
feedback algorithms. We compare the different algorithms in terms
of the number of tuning parameters and performance under both steady
and variable load. Furthermore, we identify two categories of
fairness requirements for SIP overload control, namely,
user-centric and provider-centric fairness. With the introduction
of a new double-feed SIP overload control architecture, we show
how the algorithms can meet those fairness criteria.

\end{abstract}

\section{Introduction}

The Session Initiation Protocol~\cite{RFC3261} (SIP) is a
signaling protocol standardized by IETF for creating, modifying, and
terminating sessions in
the Internet. It has been used for many session-oriented
applications, such as calls, multimedia distributions, video
conferencing, presence service and instant messaging. Major
standards bodies including 3GPP, ITU-I, and ETSI have all adopted
SIP as the core signaling protocol for Next Generation Networks
predominately based on the Internet Multimedia Subsystem (IMS)
architecture.

The widespread popularity of SIP has raised attention
to its readiness of handling overload \cite{sipping:overload}.
A SIP server can be overloaded for many reasons, such as emergency-induced
call volume, flash crowds generated by TV programs (e.g., American
Idol), special events such as ``free tickets to third caller'', or even denial
of service attacks. Although server overload is by no means
a new problem for the Internet, the key observation that
distinguishes the SIP overload problem from others is that the cost of
rejecting a SIP session usually cannot be ignored compared to the cost of
serving a session. Consequently, when a SIP server has to reject a large
amount of arriving sessions, its performance collapses. This
explains why using the built-in SIP overload control mechanism based
on generating a rejection response messages does not solve the problem.
If, as is often recommended, the rejected sessions are sent to a load-sharing
SIP server, the alternative server will soon also be generating nothing but
rejection responses, leading to a cascading failure. Another important
aspect of overload in SIP is related to SIP's multi-hop server architecture
with name-based application level routing. This aspect creates the so-called
``server to server'' overload problem that is generally not comparable
to overload in other servers such as web server.

To avoid the overloaded server ending up at a state spending all
its resources rejecting sessions, %
Hilt {\em et al.}~\cite{hilt:overload} outlined a SIP overload control
framework based on feedback from the receiving server to its upstream
sending servers. The feedback can be in terms of a
rate or a load limiting window size. However, the exact algorithms that
may be applied in this framework and the potential performance implications
are not obvious. In particular, to our best knowledge there has been
no published work on specific window-based algorithms for SIP overload
control, or comprehensive performance evaluation of rate-based feedback
algorithms that also discusses dynamic load conditions and overload
control fairness issues.

In this paper, we introduce a new dynamic session estimation
scheme which plays an essential role in applying selected control
algorithms to the SIP overload environment. We then propose three new
window-based algorithms for SIP overload. We also apply two existing
load adaption algorithms for rate-based overload control. We thus cover all
three types of feedback control mechanisms in
\cite{hilt:overload}: the absolute rate feedback, relative rate
feedback and window feedback. Our simulation evaluation results show that although the algorithms differ in their tuning parameters, most of them are able to achieve theoretical maximum performance under steady state load conditions. The results under dynamic load conditions with source arrival and departure are also encouraging. Furthermore, we look at the fairness issue in the context of SIP overload and propose the notion of user-centric fairness vs. service provider-centric fairness. We show how different algorithms may achieve the desired type of fairness. In particular, we found that the user-centric fairness is difficult to achieve in the absolute rate or window-based feedback mechanisms. We solve this problem by introducing a new double-feed SIP overload control architecture.

The rest of this paper is organized as follows:
Section~\ref{sec:background} presents background on the SIP overload problem, and discusses related work.
In Section~\ref{sec:algsip}
we propose three window-based SIP overload control algorithms and
describe two existing load adaptation algorithm to be applied for rate-based SIP overload control. Then we
present the simulation model and basic SIP overload results
without feedback control in Section~\ref{sec:simmodel}.
The steady load performance evaluation of the control algorithms are presented in Section~\ref{sec:steadyload}, followed by dynamic load performance with
fairness consideration in Section~\ref{sec:dynfair}. Finally
Section~\ref{sec:conclude} concludes the paper and discusses
future work.

\section{Background and Related Work} \label{sec:background}

\subsection{SIP Overview}

\begin{figure}
\centering
\epsfig{file=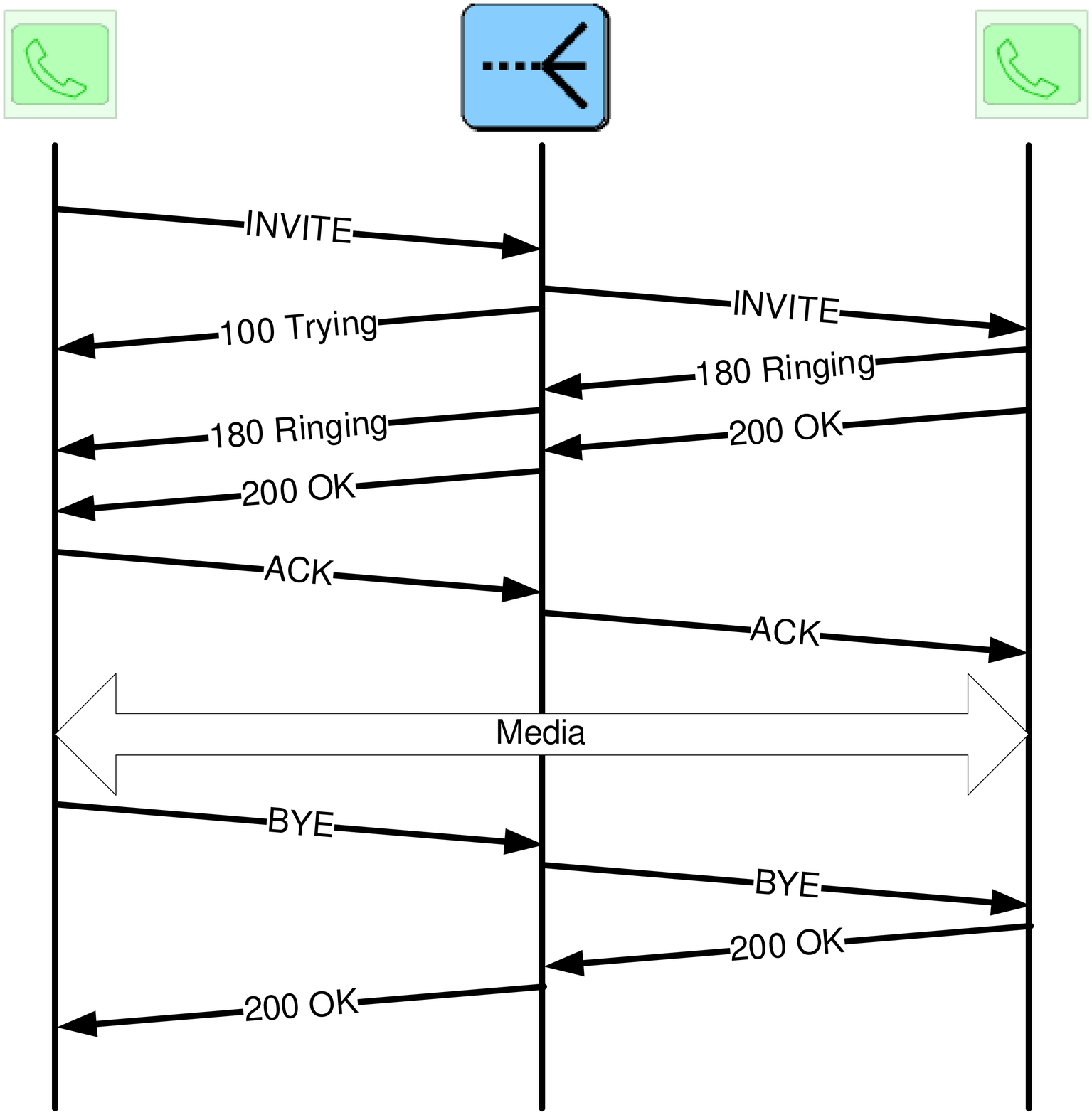,scale=0.20} \caption{SIP call session message flow} \label{fig:sigflow}
\end{figure}

SIP is a message based protocol for managing sessions. There are
two basic SIP entities, SIP User Agents (UAs), and SIP servers.
SIP servers can be further grouped into proxy servers for session
routing and registration servers for UA registration. In this paper
we focus primarily on proxy servers. In the remainder of this document,
when referring to SIP servers, we mean proxy server unless explicitly mentioned otherwise. One of the most popular session types that SIP is used for is call session. This is also the type of session we will consider in this paper. In a typical SIP call session, the caller and
callee have UA functionalities, and they set up the session through
the help of SIP servers along the path between them.
Figure~\ref{fig:sigflow} shows the SIP message flow establishing
a SIP call session. The caller starts with sending
an {\sf INVITE} request message towards the SIP proxy server,
which replies with a {\sf 100 Trying} message and forwards the
request to the next hop determined by name-based application level
routing. In Figure~\ref{fig:sigflow} the next hop for the only SIP
server is the callee, but in reality it could well be another SIP
server along the path. Once the {\sf INVITE} request finally
arrives at the callee, the callee replies with a {\sf 180 Ringing}
message indicating receipt of the call request by the callee UA, and sends
a {\sf 200 OK} message when the callee picks
up the phone. The {\sf 200 OK} message makes its way back to the caller,
who will send an {\sf ACK} message to the callee to conclude the
call setup. Afterwards, media may flow between the caller and
callee without the intervention of the SIP server. When one party wants
to tear down the call, the corresponding UA sends a {\sf BYE} message to the other
party, who will reply with a {\sf 200 OK} message to confirm the
call hang-up. %
Therefore, a typical SIP call session entails processing
of five incoming messages for call setup and two incoming messages
for call teardown, a total of seven messages for the whole
session.

SIP is an application level protocol on top of the transport
layer. It can run over any common transport layer protocol, such
as UDP and TCP. A particular aspect of SIP related to the overload
problem is its timer mechanism. SIP defines a large number of
retransmission timers to cope with message loss, especially when
the unreliable UDP transport is used. As examples, we illustrate
three of the timers which are commonly seen causing problems under
overload. The first is timer A that causes an {\sf INVITE}
retransmission upon each of its expirations. With an initial value
of $T_1 = 500~ms$, timer A increases exponentially until its total
timeout period exceeds 32~s. The second timer of
interest is the timer that controls the retransmission of
{\sf 200 OK} message as a response to an {\sf INVITE} request. The timer for {\sf 200 OK} also starts with $T_1$, and its value
doubles until it reaches $T_2 = 4~s$. At that time the timer value
remains at $T_2$ until the total timeout period exceeds 32~s. The
third timer of interest is timer E, which controls the {\sf BYE}
request retransmission. Timer E follows a timeout pattern similar to
the {\sf 200 OK} timer. Note that the receipt of
corresponding messages triggered by each of the original messages
will quench the retransmission timer. They are the {\sf 100
Trying} for {\sf INVITE}, {\sf ACK} for {\sf 200 OK}, and {\sf 200
OK} for {\sf BYE}. From this description, we know that for
example, if an {\sf INVITE} message for some reason is dropped or
stays in the server queue longer than 500~ms without generating
the {\sf 100 Trying}, the upstream SIP entity will retransmit the
original {\sf INVITE}. Similarly, if the round trip time of the
system is longer than 500~ms, then the {\sf 200 OK} timer and the
{\sf BYE} timer will fire, causing retransmission of these
messages. Under ideal network conditions without link delay and
loss, retransmissions are purely wasted messages that should be avoided.

\subsection{Types of SIP Server Overload}

There are many causes to SIP overload, but the resulting SIP overload cases can usually be grouped into either of the two types: server to server overload or client to server overload.

\begin{figure}
\centering
\epsfig{file=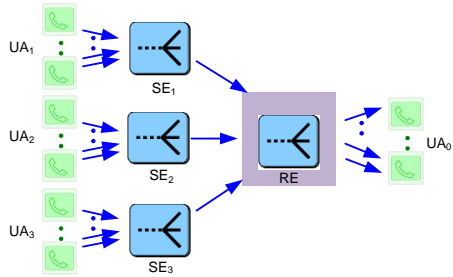, scale=0.3}
\caption{Server to server overload}
\label{fig:s2soverload}
\end{figure}

A typical server to server overload topology is illustrated in Figure~\ref{fig:s2soverload}. In this figure the
overloaded server (the Receiving Entity or $RE$) is connected with a
relatively small number of upstream servers (the Sending Entities or
$SEs$). One example of server to server overload is a special event
such as ``free tickets to the third caller'', also referred to as flash
crowds. Suppose $RE$ is the
Service Provider (SP) for a hotline N. $SE_1$, $SE_2$
and $SE_3$ are three SPs that reach the hotline through $RE$. When
the hotline is activated, $RE$ is expected to receive a large call
volume to the hotline from $SE_1$, $SE_2$ and $SE_3$ that far
exceeds its usual call volume, potentially putting $RE$ into a
severe overload.
The second type of overload, known as client-to-server overload is when a
number of clients overload the next hop server directly.  An example is avalanche restart, which happens when power is restored after a mass power failure in a large metropolitan area. At the time the power is restored, a very large number of SIP devices boot up and send out SIP registration requests almost simultaneously, which could easily overload the corresponding SIP registration server.
This paper only discusses the server-to-server overload
problem. The client-to-server overload problem may require different
solutions and is out of scope of this paper.

\subsection{Existing SIP Overload Control Mechanisms}

Without overload control, messages that cannot be processed by the
server are simply dropped.
Simple drop causes the corresponding SIP timers to fire, and further amplifies the overload situation.

SIP has a {\sf 503 Service Unavailable} response message used to reject a session request and cancel any related outstanding retransmission timers. However, because of the relatively high cost of generating this rejection, this message cannot solve the overload problem.

SIP also defines an optional parameter called ``Retry-after'' in the {\sf
503 Service Unavailable} message. The ``Retry-after'' value
specifies the amount of time that the receiving $SE$ of the message
should cease sending any requests to the $RE$. The {\sf 503 Service
Unavailable} with ``Retry-after'' represents basically an on/off
overload control approach, which is known to be unable to fully
prevent congestion collapse \cite{sipping:overload}. Another
related technique is to allow the $SE$ to fail over the rejected requests to an alternative
load-sharing server. However, in many situations the load-sharing
server could ultimately be overloaded as well, leading to
cascading failure.

\subsection{Feedback-based Overload Control}

The key to solving the SIP server overload problem is to make sure
the upstream $SEs$ only send the amount of traffic that the $RE$ is
able to handle at all times. In this ideal situation, there will
be no message retransmission due to timeout and no extra
processing cost due to rejection. The server CPU power can be
fully utilized to deliver its maximum session service capacity.

A feedback loop is a natural approach to achieve the ideal overload control goal. Through the loop, $RE$ notifies $SEs$ the amount of load that is acceptable.

\begin{figure}
\centering
\epsfig{file=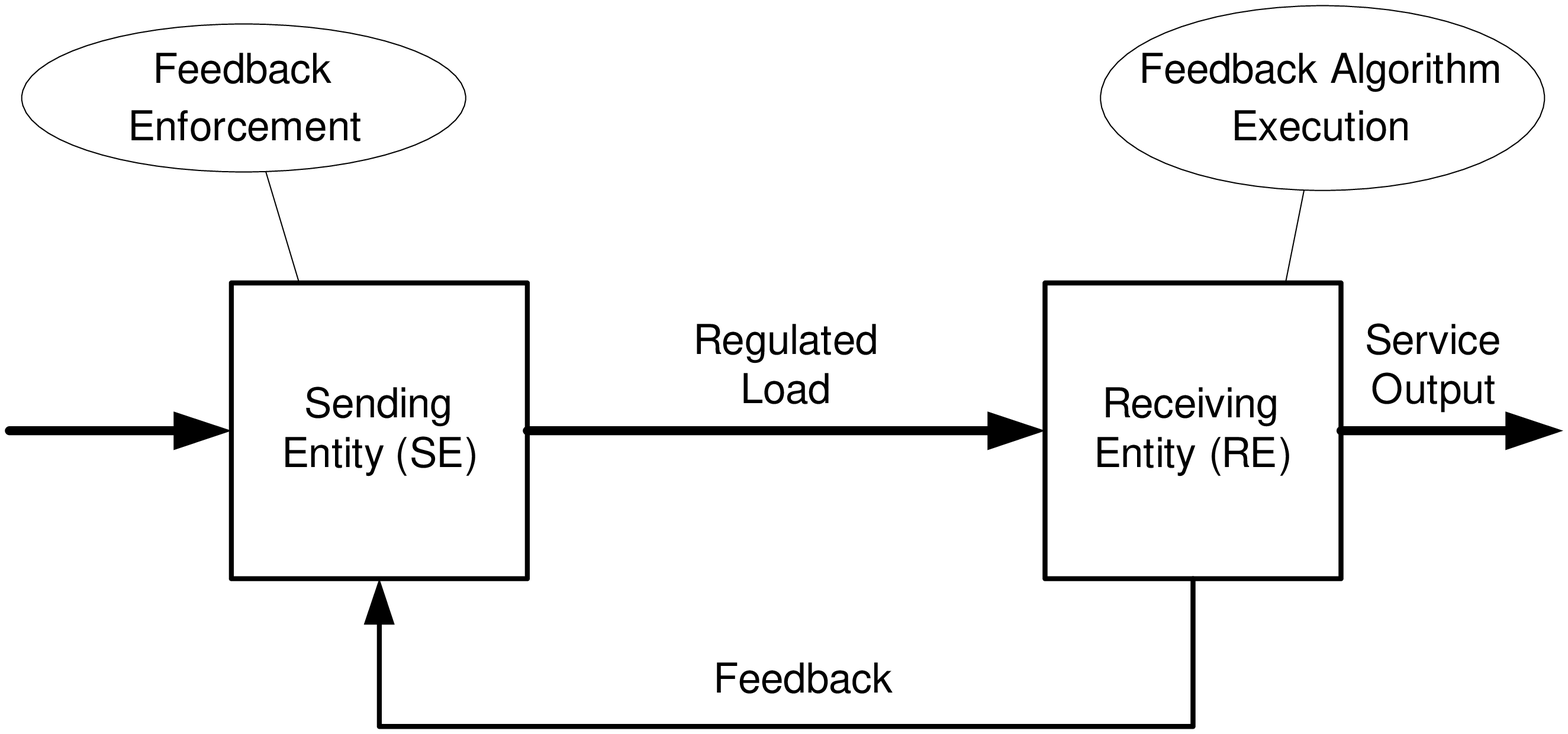, scale=0.3} \caption{Generalized feedback architecture} \label{fig:feedback}
\end{figure}

To some extent the existing SIP {\sf 503 Service Unavailable} mechanism with
the ``Retry-after'' header is a basic form of the feedback
mechanism. Unfortunately, its on/off control nature has proven to
be problematic. Therefore, the IETF community has started looking at more sophisticated
pushback mechanisms including both rate-based and window-based
feedback. A generalized model of the feedback-based control
model is shown in Figure~\ref{fig:feedback}. There are three main
components in the model: feedback algorithm execution at $RE$, feedback
communication from $RE$ to $SE$, and feedback enforcement at the $SE$.

\subsubsection{Feedback Algorithm Execution}

Absolute rate, relative rate and window feedback are three main
SIP feedback control mechanisms. Each mechanism executes specific
control algorithms to generate and adapt the feedback value.

In absolute rate-based feedback, the feedback generation
entity $RE$ needs to estimate its acceptable load and allocate it
among the $SEs$. The feedback information is an absolute load value for
the particular $SE$. The key element in absolute rate
feedback is an algorithm for dynamic acceptable load estimation.

In relative rate-based feedback, the feedback generation
entity $RE$ computes an incoming load throttle percentage based on a
target resource metric (e.g., CPU utilization).
The feedback information is a dynamic percentage value indicating
how much proportion of the load should be accepted or rejected
relative to the original incoming load. The key element in
relative rate feedback is the dynamic relative rate adjustment
algorithm and the choosing of the target metric.

In window-based feedback, the feedback generation entity
$RE$ estimates a dynamic window size for each $SE$ which specifies the
number of acceptable sessions from that particular $SE$. The
feedback information is the current window size. The key element
in window-based feedback is a dynamic window adjustment
algorithm.

The feedback generation could be either time-driven or
event-driven. In time-driven mechanisms, the control is usually
exercised every pre-scheduled control interval, while in
event-driven mechanisms, the control is executed upon the
occurrence of some event, such as a session service completion. We
will examine both time-driven and event-driven algorithms in this
paper.

\subsubsection{Feedback Enforcement Mechanisms}

The $SEs$ may choose among many well-known traffic regulation mechanisms to
enforce feedback control, such as percentage throttle, leaky
bucket and token bucket, automatic call gapping, and window
throttle. Since our focus is on the feedback
algorithms, throughout this paper we will use percentage throttle for
rate-based feedback and window-throttle for window-based feedback mechanisms.
In our percentage throttle implementation we
probabilistically block a given percentage of the load arrival
to make sure the actual output load conforms to the regulated load
value. For window throttle implementation, we only forward a
specific session arrival when there is window slot available.

\subsubsection{Feedback Communication}
The feedback information for SIP signaling overload control can be
communicated via an in-band or out-of-band channel. In this paper, we have chosen to use the in-band feedback communication approach. Specifically, any feedback
information available is sent in the next immediate message that
goes to the particular target $SE$. This approach has an advantage
in server to server overload because there is generally no
problem finding existing messages to carry feedback information under
overload and it incurs minimal overhead.

\subsection{Related Work} \label{sec:relatedwork}

Signaling overload itself is a well studied topic. Many of the
previous work on call signaling overload in general
communication networks is believed to be usable by the SIP
overload study. For instance, Hosein~\cite{atntratepatent}
presented an adaptive rate control algorithm based on estimation of
message queuing delay; Cyr {\em et al.}~\cite{cyroccpatent} described
the Occupancy Algorithm (OCC) for load balancing and overload
control mechanism in distributed processing telecommunications
systems based on server CPU occupancy; Kasera {\em et al.}~\cite{kasera01} proposed an improved OCC algorithm call
Acceptance-Rate Occupancy (ARO) by taking into consideration the
call acceptance ratio, and a Signaling RED algorithm which is a
RED variant for signaling overload control.

Specifically on SIP, Ohta~\cite{ohta06} showed through simulation the congestion collapse of SIP server under heavy load and explored the approach of using a priority queuing and Bang-Bang type of overload control. Nahum {\em et al.}~\cite{erich07} reported empirical performance results of SIP server showing the congestion collapse behavior.

In addition, Whitehead~\cite{gocap} described a unified overload control framework called GOCAP for next generation networks, which is supposed to cover SIP as well. But there has been no performance results yet and it is not clear at this time how the GOCAP framework may relate to the IETF SIP overload framework.

In the most closely related work to this paper, Noel and Johnson~\cite{noel07} presented initial results comparing a SIP network without overload control, with the built-in SIP overload control and with a rate-based overload control scheme. However, their paper does not discuss window-based control, or present performance results under dynamic load, and it does not address the overload fairness problem.

\section{Feedback Algorithms for SIP Server Overload Control} \label{sec:algsip}

The previous section has introduced the main components of SIP overload feedback control framework. In this section we investigate its key component - the feedback algorithm. We propose three window-based SIP overload control methods,
namely {\em win-disc}, {\em win-cont}, and {\em win-auto}. We also apply two existing adaptive load control algorithms for rate-based control. Before discussing algorithm details, we first introduce a dynamic SIP session estimation method which plays an important role in applying selected rate-based or window-based algorithms to SIP overload control. %

\subsection{Dynamic SIP Session Estimation} \label{sec:dynsess}

Design of SIP overload control algorithm starts with determining
the control granularity, i.e., the basic control unit. Although
SIP is a message-based protocol, different types of SIP messages
carry very different weights from admission control perspective.
For instance, in a typical call session, admitting
a new {\sf INVITE} message starts a new call and implicitly
accepts six additional messages for the rest of the session
signaling. Therefore, it is more convenient to use a SIP session
as the basic control unit.

A session oriented overload control algorithm frequently requires
session related metrics as inputs such as the session service rate. In order
to obtain session related metrics a straightforward approach is
to do a \textit{full session check}, i.e., to track the start
and end message of all SIP signaling sessions. For example, the
server may count how many sessions have been started and then
completed within a measurement interval. In the case of
a call signaling, the session is initiated by an {\sf
INVITE} request and terminated with a {\sf BYE} request.
The {\sf INVITE} and {\sf BYE} are
usually separated by a random session holding time. However, SIP
allows the {\sf BYE} request to traverse a different server from
the one for the original {\sf INVITE}. In that case, some SIP
server may only see the {\sf INVITE} request while other
servers only see the {\sf BYE} request of a signaling session.
There could also be other types of SIP signaling sessions traversing
the SIP server.%
 These factors make the applicability of the
\textit{full session check} approach complicated, if not
impossible.

We use an alternative \textit{start session check} approach to estimate SIP session service rate .
The basic idea behind is that under normal working conditions, the
actual session acceptance rate is roughly equal to the session service
rate. Therefore, we can estimate the session service rate based
only on the session start messages. Specifically, the server
counts the number of {\sf INVITE} messages that it accepts per
measurement interval $T_m$. The value of the session service rate is estimated to be \begin{math} \mu = N^{accepted}_{inv} / {T_m} \end{math}. Standard smoothing functions can be applied to the periodically measured $\mu$.

One other critical session parameter
 often needed in SIP overload control algorithms is the number of sessions remaining in the server system, assuming the server processor is preceded by a queue where jobs are waiting for service. It is very important to recognize that the number of remaining sessions is NOT equal to
the number of {\sf INVITE} messages in the queue, because the
queue is shared by all types of messages, including those {\sf non-INVITE} messages which
represent sessions that had previously been accepted into the system. All
messages should be counted for the current system backlog. Hence we
propose to estimate the current number of sessions in the queue
using Eq.~\ref{equ:sesrem}:
\vspace{-0.03in}

\begin{equation} \label{equ:sesrem}
N_{sess} = N_{inv} + \frac{N_{noninv}}{L_{sess} - 1}
\end{equation}

\vspace{-0.03in}
where $N_{inv}$ and $N_{noninv}$ are current number of {\sf INVITE} and
{\sf non-INVITE} messages in the queue, respectively. The parameter
$L_{sess}$ represents the average number of messages per-session.
$N_{inv}$ indicates the number of calls arrived at the server but yet to be
processed; $N_{noninv} / (L_{sess} - 1)$ is roughly the
number of calls already in process by the server. %

Eq.~\ref{equ:sesrem} holds for both the \textit{full
session check} and the simplified \textit{start session
check} estimation approaches. The difference is how the $L_{sess}$
parameter is obtained. When the \textit{full session check}
approach is used, the length of each individual session will be
counted by checking the start and end of each individual SIP
sessions. With our simplified \textit{start session check}
approach, the session length can be obtained by counting the
actual number of messages $N^{proc}_{msg}$, processed during the
same period the session acceptance rate is observed. The session
length is then estimated to be $L_{sess} = N^{proc}_{msg} /
N^{accepted}_{inv}$.

\subsection{Active Source Estimation} \label{sec:activese}

In some of the overload control mechanisms, the $RE$ may wish to explicitly allocate its total capacity among multiple $SEs$. A simple approach is to get the number of current active $SEs$ and divide the capacity equally. We do this by directly tracking the sources of incoming load and maintaining a table entry for each current active $SE$. Each entry has an expiration timer set to one second.

\subsection{The {\it win-disc} Window Control Algorithm}

A window feedback algorithm executed at the $RE$ dynamically computes a feedback window value for the $SE$. $SE$ will forward the load to $RE$ only if window slots are currently available. Our first window based algorithm is {\em win-disc}, the short name for {\em window-discrete}. %
The main idea is that at the end of each discrete control interval
of period $T_c$, $RE$ re-evaluate the number of new session requests
it can accept for the next control interval, making sure the
delays for processing sessions already in the server and upcoming
sessions are bounded. Assuming the $RE$ advertised window to $SE_i$
at the $k^{th}$ control interval $T^k_c$ is $w^k_i$, and the total
window size for all $SEs$ at the end of the $k^{th}$ control
interval is $w^{k+1}$, the {\em win-disc} algorithm is described
below:

\begin{flushleft}
\begin{verse}
$w^0_i := W_0$ where $W_0>0$ \\
$w^k_i := w^k_i-1$ for {\sf INVITE} received from $SE_i$ \\
$w^{k+1} := \mu^{k}{T_c} + \mu^{k}{D_B} - N^{k}_{sess}$ at the end of $T^k_c$\\
$w^{k+1}_i := round ({w^{k+1}} / {N^k_{SE}})$ \\

\end{verse}
\end{flushleft}

where $\mu^{k}$ is the current estimated session service rate. $D_B$
is a parameter that reflects the allowed budget message queuing delay. $N^{k}_{sess}$ is the estimated current
number of sessions in the system at the end of $T^k_c$. $\mu^k{T_c}$
gives the estimated number of sessions the server is able to
process in the $T^{k+1}_c$ interval. $\mu^k{D_B}$ gives the average
number of sessions that can remain in the server queue given the
budget delay. This number has to exclude the number of sessions
already backlogged in the server queue, which is $N^{k}_{sess}$.
Therefore, $w^{k+1}$ gives the estimated total number of sessions
that the server is able to accept in the next $T_c$ control
interval giving delay budget $D_B$. Both $\mu^{k}$ and
$N^{k}_{sess}$ are obtained with our dynamic session estimation
algorithm in Section~\ref{sec:dynsess}. $N^{k}_{SE}$ is the
current number of active sources discussed in
Section~\ref{sec:activese}. Note that the initial value $W_0$ is
not important as long as $W_0>0$. An example value could be
$W_0 = \mu_{eng}T_c$ where $\mu_{eng}$ is the server's engineered
session service rate.

\subsection{The {\it win-cont} Window Control Algorithm} \label{sec:wincont}

Our second window feedback algorithm is {\em win-cont}, the short
name for {\em window-continuous}. Unlike the time-driven {\em
win-disc} algorithm, {\em win-cont} is an event driven algorithm
that continuously adjusts advertised window size when the
server has room to accept new sessions. The main idea of this
algorithm is to bound the number of sessions in the server at
any time. The maximum number of sessions allowed in the server is
obtained by $N^{max}_{sess} = \mu^t{D_B}$, where $D_B$ is again the
allowed message queuing delay budget and $\mu^t$ is the
current service rate. At any time, the difference between the
maximum allowed number of sessions in the server $N^{max}_{sess}$
and the current number of sessions $N_{sess}$ is the available window to be sent as
feedback. Depending on the responsiveness requirements and
computation ability, there are different design choices. First is
how frequently $N_{sess}$ should be checked. It could be after any
message processing, or after an {\sf INVITE} message processing,
or other possibilities. The second is the threshold number of session
slots to update the feedback. There are two such thresholds, the
overall number of available slots $W_{ovth}$, and the per-$SE$
individual number of available slots $W_{indvth}$. To make the
algorithm simple, we choose per-message processing $N_{sess}$
update and fix both $W_{ovth}$ and $W_{indvth}$ to 1.
A general description of the {\em win-cont} algorithm is summarized
as below:

\begin{flushleft}
\begin{verse}
$w^0_i := W_0$ where $W_0>0$ \\
$w^t_i := w^t_i-1$ for {\sf INVITE} received from $SE_i$ \\ %
$w^t_{left} := N^{max}_{sess} - N_{sess}$ upon msg processing \\
$if (w^t_{left} \geq 1)$  \\
\hspace{15pt} $w^t_{share} = w^t_{left} / N^t_{SE}$ \\
\hspace{30pt} $w^t_{i'} := w^t_{i'} + w^t_{share}$  \\
\hspace{15pt} $if (w^t_{i'} \geq 1)$ \\
\hspace{30pt} $w^t_i := (int)w^t_{i'}$  \\
\hspace{30pt} $w^t_{i'} := (frac)w^t_{i'}$  \\

\end{verse}
\end{flushleft}

Note that since $w^t_i$ may contain a decimal part, to improve the feedback window accuracy when $w^t_i$ is small, we feedback the integer part of the current $w^t_i$ and add its decimal part to the next feedback by using a temporary parameter $w^t_{i'}$. In the algorithm description, $\mu^t$, $N_{sess}$ and $N_{SE}$ are
obtained as discussed in Section~\ref{sec:dynsess} and
Section~\ref{sec:activese}. The initial value $W_0$ is not
important and a reference value is $W_0 = \mu_{eng}T_c$ where
$\mu_{eng}$ is the server's engineered session service rate.

\subsection{The {\it win-auto} Window Control Algorithm}

Our third window feedback algorithm, {\em win-auto} stands for {\em window-autonomous}. Like
{\em win-cont} , {\em win-auto} is also an event driven algorithm.
But as the term indicates, the {\em win-auto} algorithm is able to
make window adjustment autonomously. The key design principal in
the {\em win-auto} algorithm is to automatically keep the pace of
window increase below the pace of window decrease, which
makes sure the session arrival rate does not exceed the session
service rate. The algorithm details are as follows:

\begin{flushleft}
\begin{verse}

$w^0_i := W_0$ where $W_0>0$ \\
$w^t_i := w^t_i-1$ for {\sf INVITE} received from $SE_i$ \\ %
$w^t_i := w^t_i+1$ after processing a {\em new} {\sf INVITE}
\end{verse}
\end{flushleft}

The beauty of this algorithm is its extreme simplicity. The
algorithm takes advantage of the fact that retransmission starts to occur as
the network gets congested. Then the server automatically freezes
its advertised window to allow processing of backlogged sessions
until situation improves. The only check the server does is
whether an {\sf INVITE} message is a retransmitted one or a new one, which
is just a piece of normal SIP parsing done by any existing SIP
server. There could be many variations along the same line of thinking as this algorithm, but the
one as described here appears to be one of the most natural
options.

\subsection{The {\it rate-abs} Rate Control Algorithm}

We implemented an absolute rate feedback control by applying the adaptive load algorithm of Hosein~\cite{atntratepatent}, which is also used by Noel~\cite{noel07}. The main idea is to ensure the message queuing delay does not exceed the allowed budget value. The algorithm details are as follows.

During every control interval $T_c$, the $RE$ notifies the $SE$ of the new target load, which is expressed by Eq.~\ref{equ:absrate}.

\begin{equation}\label{equ:absrate}
\lambda^{k+1} = \mu^{k} (1 - \frac{(d^k_q - D_B)}{C})
\end{equation}

where ${\mu^k}$ is the current estimated service rate and $d^k_q$ is the estimated server queuing delay at the end of the last measurement interval. It is obtained by $d^k_q = N_{sess} / \mu^{k}$, where $N_{sess}$ is the number of sessions in the server. We use our dynamic session estimation in Section~\ref{sec:dynsess} to obtain $N_{sess}$, and we refer to this absolute rate control implementation as {\em rate-abs} in the rest of this document.

\subsection{The {\it rate-occ} Rate Control Algorithm}

Our candidates of existing algorithms for relative rate based feedback control are Occupancy Algorithm (OCC) \cite{cyroccpatent}, Acceptance-Rate Occupancy (ARO), and Signaling RED (SRED)
\cite{kasera01}. We decided to implement the basic OCC algorithm because this mechanism already illustrates inherent properties with any occupancy based approach. On the other hand, tuning of RED based algorithm is known to be relatively complicated.

The OCC algorithm is based on a target processor occupancy, defined as the percentage of time the processor is busy processing messages within a measurement interval. So the target processor
occupancy is the main parameter to be specified. The processor
occupancy is measured every measurement interval $T_m$. Every
control interval $T_c$ the measured processor occupancy is
compared with the target occupancy. If the measured value is
larger than the target value, the incoming load should be
reduced. Otherwise, the incoming load should be increased. The
adjustment is reflected in a parameter $f$ which indicates the
acceptance ratio of the current incoming load. $f$ is therefore
the relative rate feedback information and is expressed by the
Eq.~\ref{equ:occ}:

\begin{equation} \label{equ:occ}
f^{k+1} = \left\{ \begin{array} {ll} f_{min}, & \textrm{if
${\phi^k}{f^k} < f_{min}$} \\ 1, & \textrm{if ${\phi^k}{f^k} > 1$} \\
{\phi^k}{f^k}, & \textrm{otherwise} \end{array} \right.
\end{equation}

where $f_k$ is the current acceptance ratio and $f_{k+1}$ is the
estimated value for the next control interval. $\phi^k =
min({\rho_B/\rho^k_t}, {\phi_{max}})$. $f_{min}$ exists to give
none-zero minimal acceptance ratio, thus prevents the server from
completely shutting off the $SE$. $\phi_{max}$ defines the maximum
multiplicative increase factor of $f$ in two consecutive control
intervals. In this paper we choose the two OCC
parameters $\phi$ and $f_{min}$ to be 5 and 0.02, respectively in
all our tests.

We will refer to this algorithm as {\em rate-occ} in the rest of this paper.

\section{Simulation Model} \label{sec:simmodel}

\subsection{Simulation Platform}

We have built a SIP simulator on the popular OPNET modeler simulation
platform~\cite{opnet}.
Our SIP simulator captures both the {\sf
INVITE} and {\sf non-INVITE} state machines as defined in
RFC3261. It is also one of the independent
implementations in the IETF SIP server overload design team, and has been calibrated in the team under common simulation scenarios.

Our general SIP server model consists of a FIFO queue followed by a SIP processor. Depending on the control mechanisms, specific
overload related pre-queue or post-queue processing may be
inserted, such as window increase and decrease mechanisms. The feedback information is included in a new {\em overload} header of each
SIP messages, and are processed along with normal SIP message
parsing. Processing of each SIP messages creates or updates
transaction states as defined by RFC3261. The transport layer is
UDP, and therefore all the various SIP timers are in effect.

Our UA model mimics an infinite number of users. Each UA
may generate calls at any rate according to a specified
distribution and may receive calls at any rate. %
The processing capacity of a UA is assumed to be infinite since we are interested in the server performance.

\subsection{Simulation Topology and Configuration}

We use the topology in Figure~\ref{fig:s2soverload} for current
evaluations. There are three UAs on the left, each of which represents
an infinite number of callers. Each UA is connected to an $SE$. The
three $SEs$ all connect to the $RE$ which is the potentially
overloaded server. The queue size is 500~messages. The core
RE connects to $UA_0$ which represents an infinite number of callees.
Calls are generated with exponential interarrival times from the callers
at the left to the callees on the right. Each call signaling
contains seven messages as illustrated in
Figure~\ref{fig:sigflow}. The call holding time is assumed to be
exponentially distributed with average of 30~seconds. The normal message processing rate and the
processing rate for rejecting a call at the $RE$ are 500 messages per second (mps) and
3000~mps, respectively.

Note that the server processor configuration, together with the call
signaling pattern, results in a nominal system service
capacity of 72~cps. All our load and goodput related values
presented below are normalized to this system capacity. Our main result metric is goodput, which counts the number of calls with successful delivery of all five call setup messages from {\sf INVITE} to {\sf ACK} below 10~s.

For the purpose of this simulation, we also made the following
assumptions. First, we do not consider any link transmission delay or loss. However, this does not mean feedback is
instantaneous, because we assume the piggyback feedback mechanism. The feedback will only be sent upon the next available
message to the particular next hop. Second, all the edge proxies are assumed to have infinite processing capacity. By removing the processing limit of the edge server, we avoid the conservative load pattern when the edge proxy server can itself be overloaded.

These simple yet classical network configuration and assumptions allow us to focus primarily on the algorithms themselves without being distracted by less important factors, which may be further explored in future work.

\vspace{-0.1in}
\begin{figure}
\centering
\epsfig{file=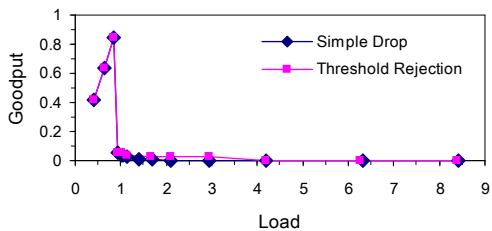} \caption{SIP overload with no feedback control} \label{fig:sim1n2}
\end{figure}

\vspace{-0.05in}

\subsection{SIP Overload Without Feedback Control}

For comparison, we first look at SIP overload performance without any feedback control. Figure~\ref{fig:sim1n2} shows the simulation results for two basic scenarios. In the ``Simple Drop'' scenario, any message arrived after the queue is full is simply dropped. In the ``Threshold Rejection'' scenario, the server compares its queue length with a high and a low threshold value. If the queue length reaches the high threshold, new {\sf INVITE} requests are rejected but other messages are still processed. The processing of new {\sf INVITE} requests will not be restored until the queue length falls below the low threshold. As we can see, the two result goodput curves almost overlap. Both cases display similar precipitous drops when the offered load approximates the server capacity, a clear sign of congestion collapse. However, the reasons for the steep collapse of the goodput are quite different in the two scenarios. In the ``Simple Drop'' case, there are still around one third of the {\sf INVITE} messages arriving at the callee, but all the {\sf 180 RINGING} messages are dropped, and most of the {\sf 200 OK} messages are also dropped due to queue overflow. In the ``Threshold Rejection'' case, none of the {\sf INVITE} messages reaches the callee, and the $RE$ is only sending rejection messages.

\section{Steady Load Performance} \label{sec:steadyload}

We summarize in Table~\ref{tab:algorithms} the parameters for all the rate-based and window-based overload control algorithms we discussed in Section~\ref{sec:algsip}. In essence, most of the algorithms have a {\em binding} parameter, three of them use the budget queuing delay $D_B$, and one uses the budget CPU occupancy $\rho_B$. All three discrete time control algorithms have a control interval parameter $T_c$.

\begin{table}
\centering
\caption{Parameter sets for overload algorithms}
\label{tab:algorithms}
\begin{tabular}{|l|l|l|l|l|}
\hline
Algorithm & Binding & Control & Measure & Additional \\
&         & Interval & Interval & \\
\hline\hline
{\em rate-abs} & $D_B$ & $T_c$ & $T_m$ &  \\
\hline
{\em rate-occ} & $\rho_B$ & $T_c$ & $T_m$ & $f_{min}$ and $\phi$  \\
\hline
{\em win-disc} & $D_B$ & $T_c$ & $T_m$ & \\
\hline
{\em win-cont} & $D_B$$^*$ & N/A & $T_m$ & \\
\hline
{\em win-auto} & N/A$^\dagger$ & N/A & N/A & \\
\hline
\end{tabular}
\begin{verse}
$D_B$: budget queuing delay \\
$\rho_B$: CPU occupancy \\
$T_c$: discrete time feedback control interval \\
$T_m$: discrete time measurement interval for selected server metrics; $T_m \leq T_c$ where applicable\\
$f_{min}$: minimal acceptance fraction \\
$\phi$: multiplicative factor \\
$^*$ $D_B$ recommended for robustness, although a fixed binding window size can also be used \\
$^\dagger$ Optionally $D_B$ may be applied for corner cases
\end{verse}
\end{table}

There is also a server metric measurement interval $T_m$ used by
four of the five algorithms. %
$T_m$ and $T_c$ need to be separate only
when $T_c$ is relatively large compared to $T_m$. The choice of the $T_m$ value
depends on how volatile the target server metric is over time. For
example, if the target metric is the server service rate, which is
relatively stable, a value of 100~ms is usually more than sufficient.
If on the other hand, the target metric is the current queue
length, then smaller or larger $T_m$
makes clear differences. In our study, when the specific algorithm
requires to measure the server service rate and CPU occupancy, we
apply $T_m$; when the algorithm requires information on the
current number of packets in the queue, we always obtain the
instant value. Our results show that $T_m = min (100~ms, T_c)$ is a
reasonable assumption, by which we basically reduce the two
interval parameters into one.

\vspace{-0.02in}
\begin{figure}
\centering
\epsfig{file=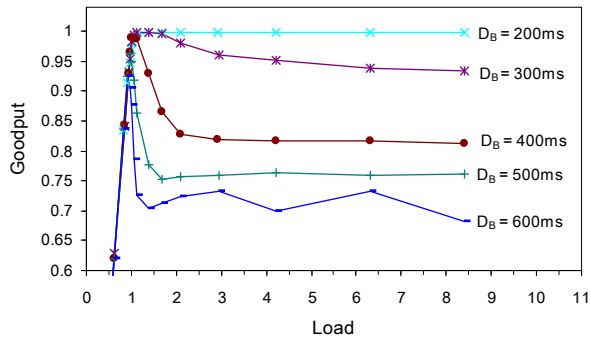} \caption{{\em win-disc} goodput under different queuing delay budget} \label{fig:windisc-bd}
\end{figure}

\begin{figure}
\centering
\epsfig{file=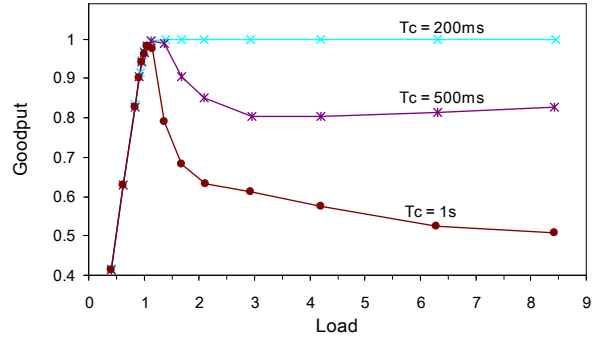} \caption{{\em win-disc} goodput under different control interval $T_c$ } \label{fig:windisc-tc}
\end{figure}

\vspace{-0.01in}
We looked at the sensitivity of $D_B$ and $T_c$ for each applicable
algorithms. Figure~\ref{fig:windisc-bd} and
Figure~\ref{fig:windisc-tc} show the results for {\em win-disc}.
All the load and goodput values have been
normalized upon the theoretical maximum capacity of the server.

We started with a $T_c$ value of 200~ms and found that the server
achieves the unit goodput when $D_B$ is set to 200~ms. Other $0 <
D_B < 200~ms$ values also showed similar results. This is not surprising
given that both the SIP caller {\sf INVITE} and callee {\sf 200
OK} timer starts at $T_1$ = 500~ms. If the queuing delay is
smaller than $1/2~T_1$ or 250~ms, then there should be no timeout
either on the caller or callee side. A larger value of $D_B$
triggers retransmission timeouts which reduce the server goodput.
For example, Figure~\ref{fig:windisc-bd} shows that at $D_B$ = 500~ms,
the goodput has already degraded by $25\%$.

Letting $D = $200~ms, we then looked at the influence of $T_c$. As
expected, the smaller the value of $T_c$ the more accurate the
control would be. In our scenario, we found that a $T_c$ value
smaller than 200~ms is sufficient to give the theatrical maximum
goodput. A larger $T_c$ quickly deteriorates the results as seen
from Figure~\ref{fig:windisc-tc}.

The effect of $D_B$ for {\em win-cont} and {\em rate-abs} show largely the similar shape, with slightly different sensitivity. Generally speaking, a positive $D_B$ value centered at around 200~ms provides a good outcome for all cases. %

\begin{figure}
\centering
\epsfig{file=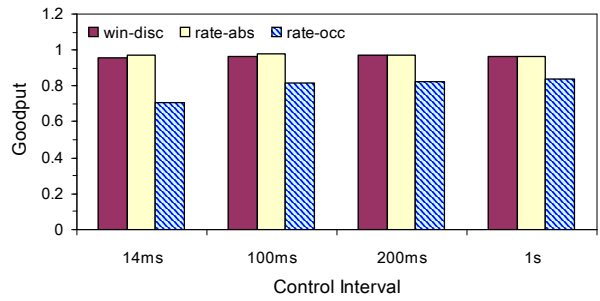} \caption{Goodput vs. $T_c$ at load 1} \label{fig:pap-compare-tc}
\end{figure}

\begin{figure}
\centering
\epsfig{file=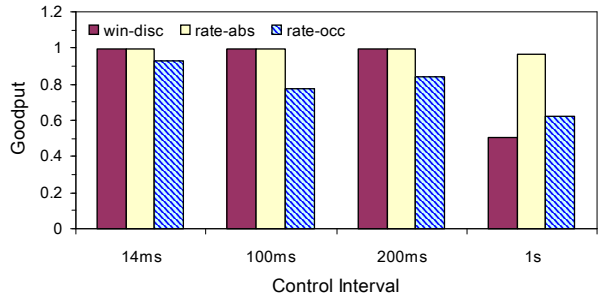} \caption{Goodput vs. $T_c$ at load 8.4} \label{fig:pap-compare-tc-L8}
\end{figure}

Figure~\ref{fig:pap-compare-tc} and
Figure~\ref{fig:pap-compare-tc-L8} compare the $T_c$ parameter for
{\em win-disc}, {\em rate-abs} and {\em rate-occ} with $D_B = 200ms$. For the {\em rate-occ} binding parameter $\rho_B$,
we used $85\%$ for the tests in Figure~\ref{fig:pap-compare-tc}
and Figure~\ref{fig:pap-compare-tc-L8}. We will explain why this
value is chosen shortly. It can be seen that the performance of
{\em win-disc} and {\em rate-abs} are very close to maximum
theoretical value in all cases except for when $T_c = 1s$ in the
heavy load case. This shows {\em win-disc} is more sensitive to
control interval than {\em rate-abs}, which could also be caused
by the more busty nature of the traffic resulted from window
throttle. It is clear that for both {\em win-disc} and {\em
rate-abs} a shorter $T_c$ improves the results, and a value below
200~ms is sufficient. Overall, {\em rate-occ} performs not as good
as the other two. But what is interesting about {\em rate-occ} is that from
14~ms to 100~ms control interval, the goodput increases in light
overload and decreases in heavy overload. This could be a result of rate
adjustment parameters which may have cut the rate too much at the
light overload.

\begin{figure}
\centering
\epsfig{file=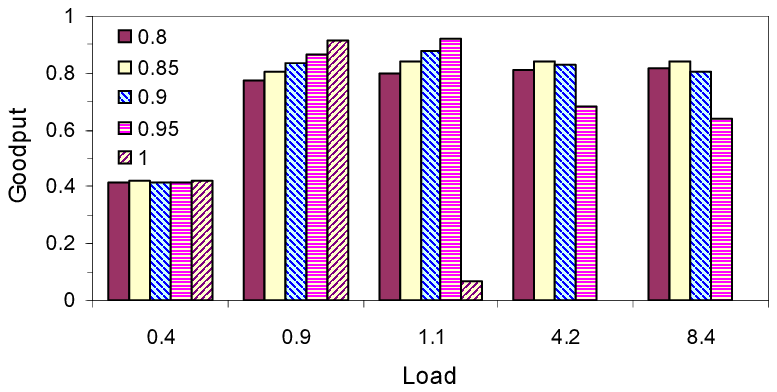} \caption{Goodput vs. $\rho_B$ at different loads} \label{fig:pap-rateocc-rho}
\end{figure}

To further understand the binding parameter $\rho_B$ of {\em rate-occ},
we illustrate in Figure~\ref{fig:pap-rateocc-rho} the relationship
between the goodput and the value of $\rho_B$ under different load
conditions. A $\rho_B$ value higher than $95\%$ easily degrades the
performance under heavy overload, because the instantaneous server
occupancy could still exceeds the healthy region and causes longer
delays which result in SIP timer expiration and message
retransmissions. A tradeoff $\rho_B$ value with the highest
and most stable performance across all load conditions in the given
scenario is $85\%$, which is the reason we used it in
Figure~\ref{fig:pap-compare-tc} and
Figure~\ref{fig:pap-compare-tc-L8}.

Finally, for the {\em win-auto} algorithm, we have found in most
cases with a reasonable initial window size in the order of 10,
the output matches perfectly the theoretical maximum line. We also
see some cases where the system could experience periods of
suboptimal yet still stable performance. The most common case
happens when the server is started with a large initial window and
the offered load is a steep jump to a heavily loaded region. Our
investigation reveals that, this suboptimal performance is caused
by the difference in the stabilized queuing delay. In most of the
normal cases, when the system reaches steady state, the queuing
delay is smaller than half of the SIP timer $T_1$ value or 250~ms.
In the suboptimal case, the system may become stable at a point
where the queuing delay can exceed 250~ms. The round-trip delay
then exceeds 500~ms, which triggers the {\sf 200 OK} timer and the
{\sf BYE} timer, each of which uses 500~ms. The two timer expirations
introduce three additional messages to the system, a
retransmitted {\sf 200 OK}, the {\sf ACK} to the retransmitted
{\sf 200 OK}, and a retransmitted {\sf BYE}. This change increases
the session length from seven to ten and reduces the maximum
server goodput by $28\%$. A cure to this situation is to introduce
an extra queuing delay parameter to the window adjustment
algorithm. Specifically, before the server increases the window
size, it checks the current queuing delay. If the queuing delay
value already exceeds the desired threshold, the window is not
increased. However, we found that determining the optimal value of the queuing
delay threshold parameter is not very straightforward and makes
the algorithm much more complex. The small chance of the occurrence of the suboptimal performance in realistic situations may not justify the additional delay binding check.

\begin{figure}
\centering \epsfig{file=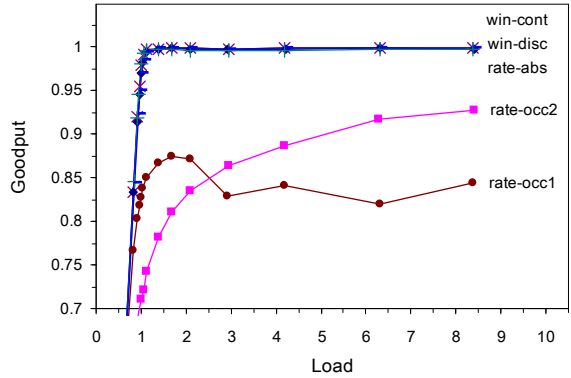} \caption{Goodput
performance for different algorithms} \label{fig:pap-compare-gp}
\end{figure}

Having looked at various parameters for all different algorithms, we now summarize the best goodput achieved by each algorithm in Figure\ref{fig:pap-compare-gp}. The specific parameters used for each algorithm is listed in Table~\ref{tab:algpram}. %

\begin{table}
\centering
\caption{Parameters used for comparison}
\label{tab:algpram}
\begin{tabular}{l|l|l|l}
\hline
& $D_B(ms)$ & $T_c(ms)$ & $T_m(ms)$  \\
\hline\hline
{\em rate-abs} & 200 & 200 & 100 \\
{\em rate-occ1$^\ddag$} & N/A & 200 & 100  \\
{\em rate-occ2$^\ddag$} & N/A & 14 & 14 \\
{\em win-disc} & 200 & 200 & 100  \\
{\em win-cont} & 200 & N/A & 100 \\
{\em win-auto} & N/A & N/A & N/A  \\
\hline
\end{tabular}
\\
$^\ddag$ in addition: $\rho_{B} = 0.85, \phi = 5, f_{min} = 0.02$
\end{table}

It is clear from Figure~\ref{fig:pap-compare-gp} that all algorithms
except for {\em rate-occ} are able to reach the theoretical
maximum goodput. The corresponding CPU occupancy
also confirms the goodput behavior. What is important to
understand is that the reason {\em rate-occ} does not operate at
the maximum theoretical goodput like the others is not simply
because of the artificial limit of setting the occupancy to
$85\%$. This point can be confirmed by the earlier
Figure~\ref{fig:pap-rateocc-rho}. The inherent issue with an
occupancy based heuristic is the fact that occupancy is not as
direct a metric as queue length or queuing delay in solving the
overload problem. Figure~\ref{fig:pap-compare-gp} shows one factor
that really helps improve the {\em rate-occ} performance at heavy
load seem to be using extremely small $T_c$. But updating the
current CPU occupancy every 14~ms is not straightforward in all
systems. Furthermore, when this short $T_c$ is used, the actual
server occupancy rose to $93\%$, which goes contrary to the original
intention of setting the $85\%$ budget server occupancy. Yet
another issue with setting the extremely short $T_c$ is its much
poorer performance than other algorithms under light overload,
which should be linked to the tuning of OCC's heuristic increase
and decrease parameters.

The merits of all the algorithms achieving maximum theoretical
goodput is that they
ensure no
retransmission ever happens, and thus the server is always busy
processing messages, with each single message being part of a
successful session.

Another metric of interest for comparison is the session setup delay, which we define as from the time the {\sf INVITE} is sent until the {\sf ACK} to {\sf 200 OK} message is received. %
We found that the {\em rate-occ} algorithm has the lowest delay but this is not significant considering it operates at the suboptimal region in terms of goodput. {\em win-cont} comes next with a delay of around 3~ms. The {\em rate-abs} offers a delay close to that of {\em win-cont} at about 3.5~ms. The remaining two {\em win-disc} and {\em win-auto} have a delay of 5~ms and 6~ms respectively. In fact all these values are sufficiently small and are not likely making any difference. %

From the steady state load analysis so far, we conclude that the
occupancy based approach is less favorable than others because of
its relatively more number of tuning parameters and not being able
to adapt to the most efficient processing condition for the
maximum goodput. {\em win-disc} and {\em abs-rate} are by
definition quite similar and they also have the same number of parameters.
Their performance is also very close, although {\em
rate-rate} has shown a slight edge, possibly because of the
smoother arrival pattern resulting from the percentage throttle. {\em
win-cont} has less tuning parameter than {\em win-disc} and {\em
abs-rate}, and offers equal or slightly better performance
Finally, {\em win-auto} is an extremely
simple algorithm yet achieves nearly perfect results in most situations.

\section{Dynamic Load Performance and Fairness for Overload Control } \label{sec:dynfair}

Although steady load performance is a good starting point for
evaluating the overload control algorithms, most of the regular
overload scenarios are not persistent steady overload. Otherwise,
The issue would become a poor capacity planning problem. The realistic server
to server overload situations are more likely short periods of
bulk loads, possibly accompanied by new sender arrivals or
departures. Therefore, in this section we extend our evaluation to
the dynamic behavior of overload control algorithms under load
variations. Furthermore, we investigate the fairness property of
each of the algorithms.

\subsection{Fairness for SIP Overload Control} \label{sec:fairness}

\begin{figure}
\centering
\epsfig{file=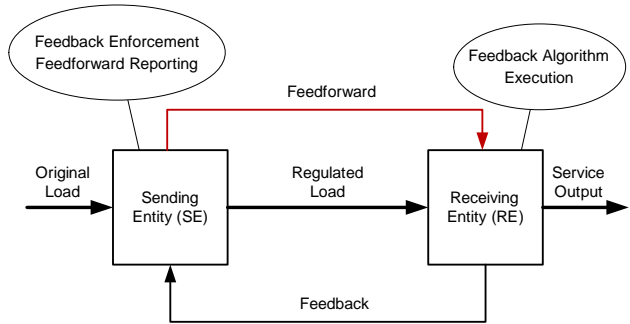, scale=0.3} \caption{The double feed architecture} \label{fig:doublefeed}
\end{figure}

\subsubsection{Defining Fairness}

Under overload, the server may allocate its available capacity
among all the upstream senders using criteria considered fair.
Theoretically, fairness can be coupled with many other factors
and could have an unlimited number of definitions.
However, we see two basic types of fairness criteria which may be applicable in most scenarios:
service provider-centric and end user-centric. %

If we consider the upstream servers representing service providers, a service-provider centric fairness means giving all the upstream servers the same aggregate success rate.

The user-centric fairness criteria aim to give each individual user who are using the overloaded server the same chance of call
success, regardless of where the call originated from. Indeed, this end user-centric fairness may be preferred in regular overload situation. For example, in the TV hotline ``free tickets to the third caller'' case, user-centric fairness ensures that all users have equal
winning probability to call in. Otherwise, a user with a
service provider who happens to have a large call volume would have
a clear disadvantage.

\subsubsection{Achieving Fairness}

Technically, achieving the basic service provider-centric
fairness is easy if the number of active sources are known, because the overloaded server simply needs to split its
processing capacity equally in the feedback generated for all the
active senders.

Achieving user-centric fairness means the overloaded server should
split is capacity proportionally among the senders based on the
senders original incoming load. For the various feedback mechanisms we
have discussed, technically the receiver in both the absolute rate-based and
window-based feedback mechanisms does not have the necessary
information to do proportional capacity allocation when the
feedback loop is activated. The receiver in the relative rate-based
mechanism does have the ability to deduce the proportion of the
original load among the senders.

To achieve user-centric fairness in absolute rate and window-based
mechanisms, we introduce a new feedforward loop in the existing
feedback architecture. The resulting double-feed architecture is
shown in Figure~\ref{fig:doublefeed}. The feedforward information
contains the sender measured value of the current incoming load.
Like the feedback, all the feedforward information is naturally
piggybacked in existing SIP messages since SIP messages by
themselves travel in both directions. This way the feedforward introduces minimal
overhead as in the feedback case. The feedforward information from
all the active senders gives the receiver global knowledge about
the original sending load. It is worth noting that, this global
knowledge equips the receiver with great flexibility that also allows
it to execute any kind of more advanced user-centric or
service provider-centric fairness criteria. Special fairness criteria may be required, for example, when the server is experiencing denial of service attack instead of regular overload. %

\begin{figure}
\centering
\epsfig{file=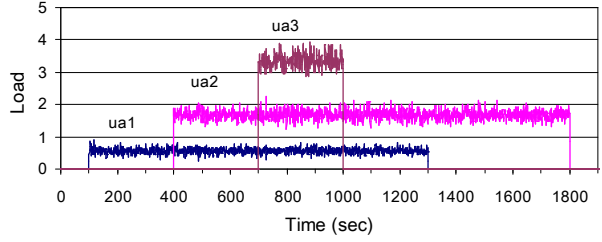} \caption{Dynamic load arrival} \label{fig:pap-dm-load}
\end{figure}

\subsection{Dynamic Load Performance}

\begin{figure}
\centering \epsfig{file=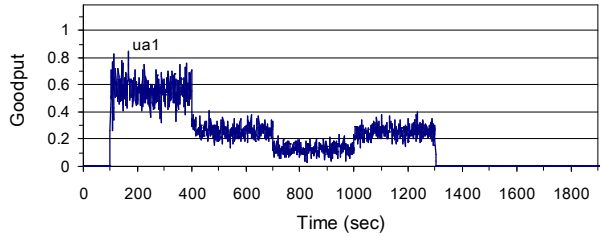}
\caption{{\em win-cont} UA1 goodput with dynamic load}
\label{fig:dm-win-cont-ua1}
\end{figure}

Figure~\ref{fig:pap-dm-load} depicts the arrival pattern for our dynamic load test. We used the step function load pattern because if the algorithm works in this extreme case, it should work in less harsh situations. The three UAs each starts and ends at different time, creating an environment of dynamic source arrival and departure. Each source also has a different peak load value, thus allowing us to observe proportional fairness mechanisms when necessary.

\begin{figure}
\centering
\epsfig{file=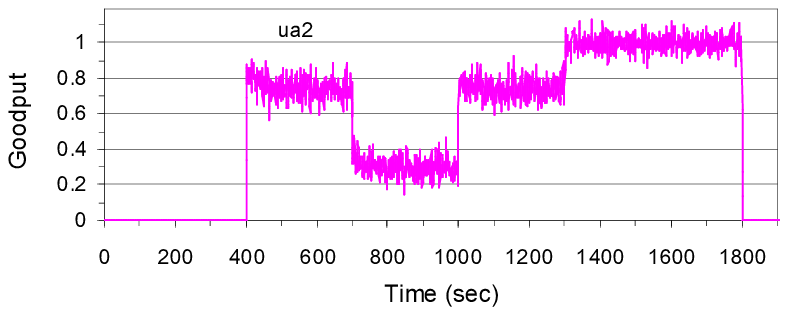} \caption{{\em
win-cont} UA2 goodput with dynamic load}
\label{fig:dm-win-cont-ua2}
\epsfig{file=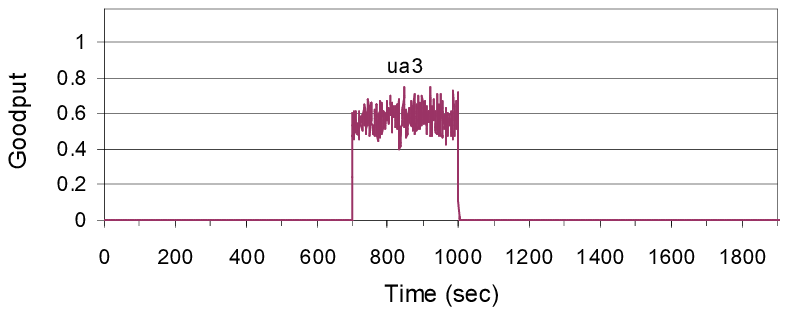} \caption{{\em win-cont} UA3 goodput with dynamic load}
\label{fig:dm-win-cont-ua3}
\end{figure}

For dynamic behavior, our simulation shows that all algorithms except {\em win-auto} adapts well to the offered dynamic load, showing little transition difference during new source arrival and existing source departure as well as at load change boundaries. As far as fairness is concerned, the {\em rate-occ} by default can provide user-centric fairness; the basic {\em rate-abs}, {\em win-disc} and {\em win-cont} algorithms are capable of basic service provider centric fairness by allocating equal amount of capacity to each $SE$. After implementing our double-feed architecture with sources reporting the original load to the $RE$, we are able to achieve user-centric fairness in all {\em rate-abs}, {\em win-disc} and {\em win-cont} algorithms through a proportional allocation of total $RE$ capacity according to $SEs$' original incoming load. In addition, having knowledge of the incoming load proportion from each $SE$ could also help us refine the algorithms when necessary. For example, in the {\em win-cont} case, we can improve the window allocation accuracy by using ``weighted fair processing'', i.e., available processing resources are probabilistically assigned to the $SEs$ based on their proportional share of total incoming load. The improved algorithm is illustrated below:

\begin{flushleft}
\begin{verse}
$w^0_i := W_0$ where $W_0>0$ \\
$w^0_{left'} := 0$ \\
$w^t_i := w^t_i-1$ for {\sf INVITE} received from $SE_i$ \\ %
$w^t_{left} := N^{max}_{sess} - N_{sess}$ upon msg processing \\
$w^t_{left} := w^t_{left} + w^t_{left'}$ \\
$if (w^t_{left} \geq 1)$  \\
\hspace{15pt} $w^t_{left'} = (frac)(w^t_{left})$ \\
\hspace{15pt} $w^t_{share} = (int)(w^t_{left})$ \\
\hspace{15pt} assuming the proportion of original load from $SE_i$ is $P\%$ \\
\hspace{30pt} $w^t_{i} := w^t_{share}$ with probability $P/100$ \\
\end{verse}
\end{flushleft}

Results of the {\em win-cont} algorithm with user-centric fairness are shown in Figure~\ref{fig:dm-win-cont-ua1} through Figure~\ref{fig:dm-win-cont-ua3}. As can be seen, UA1 starts at the 100th
second with load 0.57 and gets a goodput of the same value. At the
400th second, UA2 is started with load 1.68, three times of UA1's
load. UA1's goodput quickly declines and reaches a state where it
shares the capacity with UA2 at a one to three proportion. At the
700th second, UA3 is added with a load of 3.36. The combination
of the three active sources therefore has a load of 5.6. We see
that the goodputs of both UA1 and UA2 immediately decrease. The three
sources settle at a stable situation with roughly 0.1, 0.3, and
0.6 goodput, matching the original individual load. At the 1000th
second, the bulk arrival of UA3 ends and UA3 left the system. The
allocation split between UA1 and UA2 restores to the similar
situation before UA3's arrival at the 700th second. Finally, at
the 1300th second, UA1 departs the system, leaving UA2 with load
1.68 alone. Since the load is still over the server capacity, UA2
gets exactly the full capacity of the system with a goodput of 1.

The graph for service-provider centric fairness is similar, with
the total allocation equally shared by the current number of
active sources during each load interval.

\begin{figure}
\centering
\epsfig{file=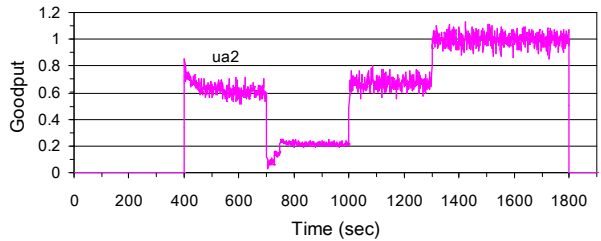}
\caption{{\em win-auto} UA2 goodput with dynamic load}
\label{fig:dm-win-auto-ua2}
\end{figure}

We also evaluated the dynamic performance of the simplest {\em
win-auto} algorithm. We found that with source arrival and
departure, the system still always reaches the maximum goodput as
long as the current load is larger than the server capacity. A
difference from the other algorithms is that it could take a
noticeably longer adaptation time to reach the steady state under
certain load surge. For example, we show in
Figure~\ref{fig:dm-win-auto-ua2} the goodput for UA2. At the 700th
second when the load increases suddenly from 2.25 to 5.6, it took
over 60~s to completely stabilize. However, the good thing
is once steady state is reached, the total goodput of all three
UAs adds up to one. Moreover, performance under source departure
is good. At the 1300th second, when UA2 becomes the only UA in the
system, its goodput quickly adapts to 1. There is, however, one
specific drawback of the {\em win-auto} mechanism. Since there is
basically no processing intervention in this algorithm, we found
it hard to enforce an explicit share of the capacity. The outcome
of the capacity split seem to be determined by the point when the
system reaches the steady state which is not easy to predict.
Therefore, {\em win-auto} may not be a good candidate when
explicit fairness is required. But because of its extreme
simplicity, as well as near perfect steady state aggregate
performance, {\em win-auto} may still be a good choice in some
situations.

\section{Conclusions and Future Work} \label{sec:conclude}

The SIP server overload problem is interesting for a number of
reasons: first, the cost of rejecting a request is not negligible
compared to the cost of serving a request; Second, the
various SIP timers lead to many retransmissions in overload and
amplify the situation; Third, SIP has a server to server
application level routing architecture. The server to server
architecture helps the deployment of a pushback SIP overload
control solution. The solution can be based on feedback of
absolute rate, relative rate, or window size.

We proposed three window adjustment algorithms {\em win-disc},
{\em win-cont} and {\em win-auto} for window-based feedback and
resorted to two existing rate adjustment algorithms for absolute
rate-based feedback {\em rate-abs} and relative rate-based
feedback {\em rate-occ}. Among these five algorithms, {\em
win-auto} is the most SIP specific, and {\em rate-occ} is the
least SIP specific. The remaining three {\em win-disc}, {\em
win-cont}, and {\em rate-abs} are generic mechanisms, and need to
be linked to SIP when being applied to the SIP environment. The
common piece that linked them to SIP is the dynamic session
estimation algorithm we introduced. It is not difficult to imagine
that with the dynamic session estimation algorithm, other generic
algorithms can also be applied to SIP.

Now we summarize various aspects of the five algorithms.

The design of most of the feedback algorithms contains a binding
parameter. Algorithms binding on queue length or queuing delay
such as {\em win-disc}, {\em win-cont} and {\em rate-abs}
outperform algorithms binding on processor occupancy such as {\em
rate-occ}. Indeed, all of {\em win-disc}, {\em win-cont} and {\em
rate-abs} are able to achieve theoretical maximum performance,
meaning the CPU is fully utilized and every message processed
contributes to a successful session, with no wasted message in the
system at all. On the other hand, occupancy based heuristic is a
much coarser control approach. The sensitivity of control also
depends on the extra multiplicative increase and decrease
parameter tuning. Therefore, from steady load performance and
parameter tuning perspective, we favor algorithms other than {\em
rate-occ}.

The adjustment performed by each algorithm can be discrete time driven such
as in {\em win-disc} and {\em rate-abs}, {\em rate-occ} or continuous
event driven such as in {\em win-cont} and {\em win-auto}. Normally
the event-driven algorithm could have smaller number of tuning
parameters and also be more accurate. But with a sufficiently
short discrete time control interval the difference between
discrete and continuous adjustments would become small.

We found that all the algorithms except {\em win-auto} adapt
well to traffic source variations as well as bulk arrival
overload. When we further look at the fairness property, especially the user-centric fairness which may be preferable in many practical situations, we found the {\em rate-occ} algorithm realizes it by default. All other algorithms except {\em win-auto} can also achieve it with our introduction of the double-feed SIP overload control architecture.

Finally, {\em win-auto} frequently needs to be singled out because
it is indeed special. With an extremely simple implementation and
virtually zero parameters, it archives a remarkable steady load
aggregate output in most cases. The tradeoff to this simplicity is
a noticeable load adaptation period upon certain load surge, and
the difficulty of enforcing explicit fairness models.

Our possible work items for the next step may include adding delay
and loss property to the link, and applying other arrival patterns as well as node failure models
to make the scenario more realistic. It would be interesting to
see whether and how the currently closely matched results of each
algorithm may differ in those situations. Another work item is
that although we currently assumed percentage-throttle for
rate-based and window-throttle for window-based control only, it
may be helpful to look at more types of feedback enforcement
methods at the $SE$ and see how different the feedback algorithms will
behave.

\section{Acknowledgments}

This project receives funding from NTT. The OPNET simulation software is donated by OPNET Technologies under its university program. Part of the research was performed when Charles Shen was an intern at IBM T.J. Watson Research Center. The authors would also like to thank Arata Koike and the IETF SIP overload design team members for helpful discussions.

\bibliographystyle{splncs}
\bibliography{rfc,i-d-self,sipol-ref}

\begin{thebibliography}{10}

\bibitem{RFC3261}
Rosenberg, J., Schulzrinne, H., Camarillo, G., Johnston, A., Peterson, J.,
  Sparks, R., Handley, M., Schooler, E.:
\newblock {SIP: Session Initiation Protocol}.
\newblock RFC 3261 (Proposed Standard) (June 2002) Updated by RFCs 3265, 3853,
  4320.

\bibitem{sipping:overload}
{J. Rosenberg}:
\newblock {Requirements for Management of Overload in the Session Initiation
  Protocol}.
\newblock Internet draft (January 2008) work in progress.

\bibitem{hilt:overload}
{V. Hilt}, {I. Widjaja}, {D. Malas}, {H. Schulzrinne}:
\newblock {Session Initiation Protocol (SIP) Overload Control}.
\newblock Internet draft (February 2008) work in progress.

\bibitem{atntratepatent}
Hosein, P.:
\newblock Adaptive rate control based on estimation of message queueing delay.
\newblock United States Patent US 6,442,139 B1 (2002)

\bibitem{cyroccpatent}
{Cyr, B.L.}, {Kaufman J.S.}, {Lee P.T.}:
\newblock Load balancing and overload control in a distributed processing
  telecommunication systems.
\newblock United States Patent US 4,974,256 (1990)

\bibitem{kasera01}
Kasera, S., Pinheiro, J., Loader, C., Karaul, M., Hari, A., LaPorta, T.:
\newblock Fast and robust signaling overload control.
\newblock Network Protocols, 2001. Ninth International Conference on (11-14
  Nov. 2001)  323--331

\bibitem{ohta06}
{M. Ohta}:
\newblock {Overload Protection in a SIP Signaling Network}.
\newblock In: {International Conference on Internet Surveillance and Protection
  (ICISP'06))}. (2006)

\bibitem{erich07}
{Erich Nahum and John Tracey and Charles Wright}:
\newblock {Evaluating SIP server performance}.
\newblock In: {ACM SIGMETRICS Performance Evaluation Review}. Volume~35. (June
  2007)  349--350

\bibitem{gocap}
Whitehead, M.:
\newblock {GOCAP} - one standardised overload control for next generation
  networks.
\newblock BT Technology Journal \textbf{23}(1) (2005)  144--153

\bibitem{noel07}
Noel, E., Johnson, C.:
\newblock Initial simulation results that analyze {SIP} based {VoIP} networks
  under overload.
\newblock In: ITC. (2007)  54--64

\bibitem{opnet}
OPNET:
\newblock http://www.opnet.com

\end{thebibliography}
\end{document}